\documentclass[twocolumn,prd,amsmath,superscriptaddress]{revtex4}
\usepackage{graphicx}
\usepackage{latexsym}
\usepackage{amsmath}
\usepackage{amssymb}
\usepackage{graphics}
\usepackage{color}
\usepackage{ifthen}
\usepackage{bm}
\usepackage{color}

\definecolor{darkblue}{rgb}{0.15,0.35,0.55}
\definecolor{orangeish}{rgb}{0.65, 0.2, 0.2}
\usepackage[linktocpage=true]{hyperref}
\hypersetup{
colorlinks=true,
citecolor=darkblue,
linkcolor=orangeish,
urlcolor=darkblue,
pdfauthor={},
pdftitle={},
pdfsubject={hep-th}
}

\newcommand\ee{\end{equation}}
\newcommand\be{\begin{equation}}
\newcommand\eea{\end{eqnarray}}
\newcommand\bea{\begin{eqnarray}}

\def\gsim{ \lower .75ex \hbox{$\sim$} \llap{\raise .27ex \hbox{$>$}} }

\def\bea{\begin{eqnarray}}
\def\eea{\end{eqnarray}}

\newcommand{\rd}{{\rm d}}

\begin{document}

\title{A Positive Energy Theorem for $P(X, \phi)$ Theories}

\author{Benjamin Elder}
\affiliation{Center for Particle Cosmology, University of Pennsylvania, Philadelphia, PA 19104, USA 
}%

\author{Austin Joyce}
\affiliation{Enrico Fermi Institute and Kavli Institute for Cosmological Physics, University of Chicago, Chicago, IL 60637
}%

\author{Justin Khoury}
\affiliation{Center for Particle Cosmology, University of Pennsylvania, Philadelphia, PA 19104, USA 
}%

\author{Andrew J. Tolley}
\affiliation{CERCA/Department of Physics, Case Western Reserve University, 10900 Euclid Ave, Cleveland, OH 44106, USA
}%

\begin{abstract}
\noindent
We describe a positive energy theorem for Einstein gravity coupled to scalar fields with first-derivative interactions, so-called $P(X, \phi)$ theories.
We offer two independent derivations of this result. The first method introduces an auxiliary field to map the theory to a lagrangian describing two canonical scalar fields,
where one can apply a positive energy result of Boucher and Townsend. The second method works directly at the $P(X, \phi)$ level and uses spinorial arguments introduced by Witten. The latter approach
follows that of {\tt arXiv:1310.1663}, but the end result is less restrictive. We point to the technical step where our derivation deviates from theirs.
One of the more interesting implications of our analysis is to show it is possible to have positive energy in cases where dispersion relations following from locality and S-Matrix analyticity are violated.
\end{abstract}

\maketitle

In recent years there has been much interest in derivatively coupled scalar theories, particularly in cosmology, but also in other areas of high-energy physics and condensed matter. The novelty of these theories is that, in certain cases, they can have large classical non-linearities while remaining radiatively stable, allowing for a range of interesting phenomena. Ghost condensation~\cite{ArkaniHamed:2003uy} and galileons~\cite{Nicolis:2008in} possess time-dependent solutions that can violate the Null Energy Condition (NEC)~\cite{Creminelli:2006xe,Nicolis:2009qm,Hinterbichler:2012yn,Rubakov:2013kaa,Elder:2013gya} and yield novel cosmologies~\cite{Buchbinder:2007ad,Creminelli:2007aq,Creminelli:2010ba,Lin:2010pf,Creminelli:2012my,Hinterbichler:2012fr}. These examples are free of ghost or gradient instabilities, but have other unwelcome features, such as superluminality and conflict with black hole thermodynamics~\cite{Dubovsky:2006vk}, casting doubt on whether they admit a local ultraviolet (UV) completion~\cite{Adams:2006sv}.

It is natural to wonder if there are any statements one can make about the viability of these theories in the presence of gravity. One desirable property is that the vacuum be classically stable. This will be the case if the the theory admits a positive energy theorem for asymptotically flat solutions, {\it i.e.}, the ADM mass is always non-negative and is zero for Minkowski space only~\cite{ftnote1}.  It was originally shown~\cite{Schon:1979rg} that Einstein gravity plus matter has positive energy if the matter obeys the dominant energy condition (DEC)~\cite{ftnote2}. This proof was later simplified using a spinor technique due to Witten~\cite{Witten:1981mf, Parker:1981uy,Nester:1982tr}. (Similar proofs exist for asymptotically anti-de Sitter~\cite{Breitenlohner:1982bm, Abbott:1981ff,Gibbons:1982jg, Mezincescu:1984ev} and de Sitter~\cite{Shiromizu:2001bg,Kastor:2002fu} spacetimes.)  The result was extended by Boucher and Townsend, who showed that the DEC is not necessary to ensure positive energy~\cite{Townsend:1984iu,Boucher:1984yx}. See also~\cite{Deser:2006gt}. For a nonlinear $\sigma$-model with $N$ scalars,
\be
{\cal L}  = - \frac{1}{2}f_{IJ}(\phi)\partial_\mu\phi^I\partial^\mu\phi^J -V(\phi^I)~,
\label{nlsmaction}
\ee
where $f_{IJ}$ is positive-definite, positivity is guaranteed so long as $V(\phi^I)$ is derivable from a ``superpotential" $W(\phi^I)$ obeying the equation~\cite{ftnote3}:
\be
V(\phi^I) = 8f^{IJ}W_{,\phi^I} W_{,\phi^J}-12W^2~,
\label{superpotential}
\ee
assuming that $V(\phi^I)$ admits a minimum with $V(\bar\phi^I) \leq 0$.

In this Letter, we further extend this result and derive a positive energy theorem for scalar theories of the form 
\be
{\cal L} = P(X, \phi)~,
\label{PXaction}
\ee
by similarly constraining the functional form that $P(X, \phi)$ can take. Here $X$ is the canonical kinetic term: $X = -\frac{1}{2} (\partial \phi)^2$. (We use the mostly-plus sign convention.) This class of theories has a long history, especially in cosmology. They can be used for inflation~\cite{ArmendarizPicon:1999rj,Silverstein:2003hf,Mukhanov:2005bu}, dark energy~\cite{ArmendarizPicon:2000dh,ArmendarizPicon:2000ah}, bouncing cosmologies~\cite{Buchbinder:2007ad,Creminelli:2007aq,Lin:2010pf}, and display screening around heavy sources~\cite{Babichev:2009ee,Dvali:2010jz,Brax:2012jr,Burrage:2014uwa,deRham:2014wfa}. 

We establish the positive energy result in two different ways. First, at the classical level we map~\eqref{PXaction} to an equivalent two-derivative theory
via an auxiliary field~\cite{Tolley:2009fg}. Turning on a small kinetic term for this second field, the action takes the form~\eqref{nlsmaction}. We can then apply the result~\eqref{superpotential}, which is translated to a statement about $P(X,\phi)$ upon integrating out the auxiliary field. 

Second, we will reproduce this result directly at the $P(X,\phi)$ level using Witten's spinor arguments. This approach was taken in~\cite{Nozawa:2013maa}, although we will see that their result was slightly too restrictive. We will show that relaxing a small technical assumption in their argument allows for greater flexibility in choosing the functional form of $P(X, \phi)$.

This broader assortment of $P(X,\phi)$ theories consistent with positive energy allows for interesting phenomena. 
In particular, consider $P(X) = X + \alpha X^2$, arguably the simplest $P(X,\phi)$ example. With $\alpha > 0$, this theory obeys the DEC and hence has positive energy.
Even with $\alpha < 0$, however, we will show that the theory allows positive energy, as long as we restrict to the region $P_{,X} > 0$. This is remarkable since this theory 
with $\alpha < 0$ both exhibits a screening mechanism and violates some of the S-matrix analyticity requirements for a local theory~\cite{Adams:2006sv}. 

\vspace{0.12cm}
\noindent {\it Two-Field Description}: A $P(X,\phi)$ theory can be mapped to a 2-derivative action by introducing an auxiliary field $\chi$~\cite{Tolley:2009fg} so that the lagrangian takes the form
\be
{\cal L} = - \frac{1}{2} P_{,\chi} (\partial \phi)^2 - \chi P_{,\chi} + P\,,
\label{interm}
\ee
where $P = P(\chi, \phi)$. Indeed, the equation of motion for $\chi$ is $P_{,\chi\chi} (X - \chi) = 0$, which sets $\chi = X$, as long as $P_{,\chi\chi}\neq 0$.
Substituting $\chi = X$ in~\eqref{interm} gives ${\cal L} = P(X, \phi)$, establishing the classical equivalence of the two descriptions.
To put it in the form~\eqref{nlsmaction}, we simply turn on a small kinetic term for $\chi$:
\be
{\cal L} = - \frac{1}{2} P_{,\chi} (\partial \phi)^2 - \frac{1}{2} Z^2 (\partial \chi)^2 - \chi P_{,\chi} + P~.
\label{2fieldaction}
\ee
At this level, this is just a technical trick --- at the end we will take $Z \to 0$. Upon making the identifications
\be
f_{\chi\chi} = Z^2~;~~~~~f_{\phi\phi} =  P_{,\chi}~;~~~~~V(\chi, \phi) = \chi P_{,\chi} - P\,,
\label{comps}
\ee
this is of form~\eqref{nlsmaction}.
Note $f_{IJ}$ must be positive-definite, imposing $P_{,\chi} > 0$. After integrating out $\chi$, this translates to
$P_{,X} > 0$,
which is equivalent to the NEC~\cite{ftnote4}. In some cases this will restrict the range of $X$, but this is acceptable because it is a {\it Lorentz-invariant} restriction on the space of allowed solutions. 
The condition $P_{,X}>0$ is required for the validity of the single-field EFT which is partially UV completed by the two-field system (\ref{2fieldaction}) \cite{Tolley:2009fg}.

Substituting~\eqref{comps}, the condition~\eqref{superpotential} yields
\be
\chi P_{,\chi} - P = 8 \frac{W_{,\phi}^2}{P_{,\chi}} + 8 \frac{W_{,\chi}^2}{Z^2} - 12 W^2~.
\label{PXmastereq0}
\ee
To have a smooth $Z \to 0$ limit, the superpotential must take the form
$
W(\chi, \phi) = {\cal W}(\phi) + \frac{Z}{2\sqrt{2}} {\cal G}(\chi, \phi) + {\cal O}(Z^2)~,
$
where the factor of $2\sqrt{2}$ is introduced to simplify later expressions.
Substituting this into~\eqref{PXmastereq0} and taking $\chi \rightarrow X$,
the positive energy condition becomes
\be
P- XP_{,X}+ 8\frac{{\cal W}^2_{,\phi}}{P_{,X}}+{\cal G}^2_{,X} - 12 {\cal W}^2 = 0~.
\label{PXmastereq}
\ee
This is our main result. It is the analogue of~\eqref{superpotential} for theories of the $P(\phi,X)$ type.
Positivity of the energy requires the existence of two functions, ${\cal W}(\phi)$ and ${\cal G}(\phi,X)$,
related to $P(\phi,X)$ through~\eqref{PXmastereq}. Asymptotically, we assume $X\rightarrow 0$ and
$\phi \rightarrow \phi_0$ such that $P_{,\phi} (\phi_0)  = 0$.

The proof generalizes to $N$ scalar fields with $P(X^{I J}, \phi^K)$, where following~\cite{Nozawa:2013maa} we have defined the tensor
$X^{I J} = -\frac{1}{2} \partial_\mu \phi^I \partial^\mu \phi^J$.
This generalization is particularly interesting because the EFT of fluids~\cite{Dubovsky:2011sj} is a theory of this type.
We introduce a matrix of scalar fields $\chi^{I J}$, and the generalization of \eqref{2fieldaction} becomes
\bea \nonumber
{\cal L} = - \frac{1}{2} P_{M N} \partial^\mu \phi^M \partial_\nu \phi^N  &-& \frac{1}{2} Z^2 P_{K M} P_{L N} \partial_\mu \chi^{K L} \partial_\nu \chi^{M N} \\ 
&+& P - \chi^{M N} P_{M N}~,
\eea
where $P_{I J} \equiv \partial P / \partial \chi^{I J}$ is positive definite and invertible.  Again, integrating out $\chi$ and setting $Z \to 0$ gives $X^{I J} = \chi^{I J}$.  Following the same steps as before, we find that the superpotential must take the form
$
W = {\cal W}(\phi^I) + \frac{Z}{2 \sqrt{2}} {\cal G}(\phi^I, \chi^{M N}) + {\cal O}(Z^2)~.
$
Writing the inverse of $P_{I J}$ as $P^{I J}$, we arrive at the positivity condition
\bea \nonumber
P &-& X^{M N} P_{M N} + 8 P^{M N} {\cal W}_{, \phi^M} {\cal W}_{, \phi^N} \\
 &+& P^{K M} P^{L N} {\cal G}_{K L} {\cal G}_{M N} - 12 {\cal W}^2 = 0~.
\eea

\vspace{0.12cm}
\noindent {\it Direct derivation}: We now re-derive the positive energy condition~\eqref{PXmastereq} directly at the level of $P(X,\phi)$. This method generally follows the presentation of Witten's proof of the positive energy theorem in~\cite{Nozawa:2013maa}, but with a crucial difference, which we will point out below.

The starting point is the Nester 2-form~\cite{Witten:1981mf,Nester:1982tr}:
\be
N^{\mu\nu} = -i\left(\bar\epsilon \gamma^{\mu\nu\rho}\hat{\nabla}_\rho\epsilon-\overline{\hat{\nabla}_\rho\epsilon}\gamma^{\mu\nu\rho}\epsilon\right)~.
\label{nester2form}
\ee
where we have defined the super-covariant derivative
\be
\hat\nabla_\mu\epsilon = \left(\nabla_\mu+{\cal A}_\mu\right)\epsilon~.
\ee
Some words on notation: $\epsilon$ is a {\it commuting} Dirac spinor~\cite{Townsend:1984iu}, with conjugate $\bar\epsilon = i\epsilon^\dagger\gamma^0$; the Dirac matrices obey the Clifford algebra
$\left\{\gamma_\mu, \gamma_\nu\right\} = 2 g_{\mu\nu}$, and we have defined the anti-symmetric product $\gamma^{\mu\nu\rho} \equiv \gamma^{[\mu}\gamma^\nu\gamma^{\rho]}$.

The virtue of $N^{\mu\nu}$ is that its integral is simply related to the energy of a gravitating system~\cite{Witten:1981mf,Nester:1982tr, Townsend:1984iu}
\be
E = \int_{\partial\Sigma} \rd\Sigma_{\mu\nu} N^{\mu\nu} = \int_\Sigma \rd\Sigma_\nu\nabla_\mu N^{\mu\nu}\,,
\label{Eeqn}
\ee
where $\Sigma$ is an arbitrary space-like surface, with $\rd\Sigma_\nu$ denoting the normal-pointing volume form.
The divergence of $N^{\mu\nu}$ is given by~\cite{Nozawa:2013maa}
\be
\nabla_\nu N^{\mu\nu} = 2i \overline{\hat\nabla_\nu\epsilon}\gamma^{\mu\nu\rho}\hat\nabla_\rho\epsilon - \frac{T_{~\nu}^\mu}{M_{\rm Pl}^2}\, i\bar\epsilon\gamma^\mu\epsilon -i\bar\epsilon\gamma^{\mu\nu\rho}{\cal F}_{\nu\rho}\epsilon~,
\label{PXnester}
\ee
where ${\cal F}_{\nu\rho} = \nabla_\nu{\cal A}_\rho-\nabla_\rho{\cal A}_\nu+\left[{\cal A}_\nu, {\cal A}_\rho\right]$ is the curvature of the connection ${\cal A}_\mu$.
The stress tensor for~\eqref{PXaction} is
\be
T_{\mu\nu} = P_{,X}\partial_\mu\phi\partial_\nu\phi+Pg_{\mu\nu}~.
\label{TPX}
\ee

The term $2i \overline{\hat\nabla_\nu\epsilon}\gamma^{\mu\nu\rho}\hat\nabla_\rho\epsilon$, gives a positive contribution to the energy, after imposing the Witten condition $\gamma^i\hat{\nabla}_i\epsilon=0$~\cite{Witten:1981mf}.
The other two terms are not manifestly positive. To proceed, we follow~\cite{Nozawa:2013maa} and make the ansatz
\be
{\cal A}_\mu = {\cal W}(\phi)\gamma_\mu\,,
\ee
for some ${\cal W}(\phi)$. The last term in~\eqref{PXnester} becomes
\be
-i \bar\epsilon \gamma^{\mu\nu\rho} {\cal F}_{\nu\rho}\epsilon = - 4i\bar\epsilon\gamma^{\mu\nu}\epsilon{\cal W}_{,\phi}\partial_{\nu}\phi + 12i\bar\epsilon\gamma^\mu\epsilon {\cal W}^2~.
\ee
Our goal is to write this as a sum of squares of spinors, plus a remainder piece. To do this, we define
\begin{align}
\delta\lambda_1 &= \frac{1}{\sqrt{2}}\bigg(\sqrt{P_{,X}}\gamma^\mu\partial_\mu\phi-4\frac{{\cal W}_{,\phi}}{\sqrt{P_{,X}}}\bigg)\epsilon~; \nonumber \\
\delta\lambda_2 &=  {\cal G}_{,X} \epsilon~,
\label{spinordefs}
\end{align}
so that 
\begin{align}
\label{Fmunuintermediate} 
 \nonumber 
 &  -i \bar\epsilon \gamma^{\mu\nu\rho}{\cal F}_{\nu\rho}\epsilon = i\sum_{i=1}^2 \overline{\delta\lambda}_i \gamma^\mu \delta\lambda_i + i\bar\epsilon\gamma^\nu\epsilon P_{,X}\partial^\mu\phi\partial_\nu \phi\\ 
 &  ~~~~~~~~ + i\bar\epsilon\gamma^\mu\epsilon \bigg( XP_{,X}-8\frac{{\cal W}^2_{,\phi}}{P_{,X}}- {\cal G}_{,X}^2 +12{\cal W}^2 \bigg)~.
\end{align}
This is the key difference from the derivation in~\cite{Nozawa:2013maa}. In that calculation, the authors only used one $\delta\lambda$ spinor field, which led to a restricted class of solutions. 
Instead we expressed $-i \bar\epsilon \gamma^{\mu\nu\rho}{\cal F}_{\nu\rho}\epsilon$ as the sum of {\it two} squares of spinors. The second spinor introduces a new function ${\cal G} = {\cal G}(X, \phi)$,
which allows us to derive a more general positivity constraint than~\cite{Nozawa:2013maa}.

Combining~\eqref{PXnester},~\eqref{TPX} and~\eqref{Fmunuintermediate}, we obtain
\begin{align}
\nonumber
&  \nabla_\nu N^{\mu\nu} = 2i \overline{\hat\nabla_\nu\epsilon}\gamma^{\mu\nu\rho}\hat\nabla_\rho\epsilon + i\sum_{i=1}^2 \overline{\delta\lambda}_i \gamma^\mu \delta\lambda_i \\
& + i\bar\epsilon\gamma^\mu\epsilon\bigg(XP_{,X}-P-8\frac{{\cal W}^2_{,\phi}}{P_{,X}}-{\cal G},_X^2+12{\cal W}^2\bigg)~.
\label{nesterdiv}
\end{align}
The first line is positive-definite, whereas the second line is not. To ensure positivity of $E$, it is sufficient to set the second line to zero. This yields~\eqref{PXmastereq}, which is precisely the energy condition obtained from the 2-field approach. The mass vanishes for $\hat{\nabla}_\mu \epsilon = \delta\lambda_a = 0$, which implies Minkowski or AdS space-time~\cite{Townsend:1984iu}.
Having derived this constraint on the functional form of $P$, we now turn to solving this equation in a few situations of interest~\cite{ftnote5}.

\vspace{0.12cm}
\noindent {\it Pure $P(X)$}: One simple but nontrivial case to consider is $P = P(X)$, {\it i.e.}, a field with purely derivative couplings and no potential. We simply assume that ${\cal W} \equiv {\cal W}_0$ is constant, and take ${\cal G} = {\cal G}(X)$. In this case, the positive energy condition~\eqref{PXmastereq} reduces to an ordinary differential equation for ${\cal G}$, which can be integrated:
\be
{\cal G}(X) = \int {\rm d}X \left( XP_{,X} - P + 12{\cal W}_0^2\right)^{1/2} \,.
\label{purePXpositivity}
\ee
In order for this integral to be real-valued, we must have $X P_{,X} - P \geq -12{\cal W}_0^2$. Note that this condition is weaker than the dominant energy condition: $X P_{,X} - P \geq 0$.

As a simple example, consider the function
\be
P(X) = X - \beta X^2~; \qquad \beta \geq 0\,.
\label{betaeg}
\ee
This theory violates the DEC for all $X$: $X P_{,X} - P  = - \beta X^2 < 0$. Recall that our derivation requires $P_{,X} \geq 0$, so we must restrict
ourselves to the range $\lvert X\rvert \leq 1/\sqrt{2}\beta$. In this case,~\eqref{purePXpositivity} can be integrated, ensuring the existence of a suitable superpotential, and guaranteeing that the theory has positive energy in the allowed $X$ range.

This theory with ``wrong-sign" $X^2$ term is well-known to violate the standard dispersion relations following from local S-matrix theory~\cite{Adams:2006sv}, at least at tree level. 
Nevertheless, we have shown that the theory does allow positive energy, at least over the range of $X$ where the NEC is satisfied. This may seem
paradoxical from the perspective of the 2-field action discussed earlier; after all,~\eqref{2fieldaction} describes two healthy scalars with some potential, and
therefore should have an analytic S-matrix. The resolution is that the vacuum state $X = 0$ or, equivalently, $\chi = 0$, is tachyonic in the two-field language,
hence its S-matrix is ill-defined.

\vspace{0.12cm}
\noindent {\it Separable $P(X, \phi)$}: A slightly more complicated case is where $P$ is a separable function:
\be
P(X, \phi) = K(\phi) \tilde P(X) - V(\phi)~,
\ee
with $ K(\phi)\geq 0$ without loss of generality. This form has been widely-studied in the context of k-essence~\cite{ArmendarizPicon:1999rj,ArmendarizPicon:2000dh}. 

It will prove convenient to redefine the arbitary function ${\cal G}(X,\phi)$ via
\be
{\cal G}_{,X}^2  = {\cal H}(X, \phi) + 8 \frac{{\cal W}_{,\phi}^2}{K(\phi)} \bigg(1 - \frac{1}{\tilde{P}_{,X}}\bigg)\,.
\label{Gredef}
\ee
Inserting this into~\eqref{PXmastereq}, we find that $P$ must satisfy
\be
\nonumber
\tilde P - X \tilde P_{,X} + \frac{{\cal H}(X, \phi)}{K(\phi)} = \frac{1}{K(\phi)}\bigg(12 {\cal W}^2 - 8 \frac{{\cal W},_\phi^2}{K} + V(\phi)\bigg)\,.
\ee
For this to be separable, ${\cal H}$ must factorize as
${\cal H}(X, \phi) = K(\phi) H(X)$.
The above then implies two equations
\bea 
\nonumber
H(X) &=& X \tilde P_{,X} - \tilde P(X) - E \,;\\
V(\phi) &=& 8 \frac{{\cal W},_\phi^2}{K(\phi)} - 12 {\cal W}^2 + E K(\phi)~.
\label{separableP}
\eea

We must ensure that through all these redefinitions we maintain ${\cal G},_X^2 \geq 0$. Combining~\eqref{Gredef}--\eqref{separableP},
we find 
\be
X \tilde P_{,X} - \tilde P(X) \geq E -  8 \frac{{\cal W}_{,\phi}^2}{K^2(\phi)} \bigg(1 - \frac{1}{\tilde{P}_{,X}}\bigg)\,.
\ee
This allows for DEC-violation through the kinetic part of the action whenever the right-hand side is negative.

A few limiting cases of these results:
\begin{itemize}

\item If $\tilde{P} = X$, corresponding to the two-derivative lagrangian ${\cal L} = K(\phi) X - V(\phi)$, we can set $E= 0$ and ${\cal G} = 0$.
The second of~\eqref{separableP} reduces to the standard result~\eqref{superpotential} for a single scalar field
\be
V(\phi) = 8 \frac{{\cal W},_\phi^2}{K(\phi)} - 12 {\cal W}^2 \,.
\ee

\item For the pure $P(X)$ case, corresponding to $K(\phi)  = 1$ and $V(\phi) = 0$, the second of~\eqref{separableP} allows us to choose ${\cal W} = {\cal W}_0 = {\rm constant}$,
with $E = 12 {\cal W}_0^2$. The first of~\eqref{separableP}, combined with~\eqref{Gredef}, then implies
\be
{\cal G}_{,X}^2  = {\cal H}(X) = XP_{,X} - P + 12{\cal W}_0^2\,,
\ee
whose integral reproduces~\eqref{purePXpositivity}.

\end{itemize}

\noindent {\it Conclusions}: We derived, following two different methods, an extension of the positive energy theorem of General Relativity to the class of $P(X, \phi)$ scalar field theories. 
We found that as long as it is possible to write $P$ in terms of two arbitrary superpotential-like functions, positive energy is guaranteed. This derivation generalizes the
result of~\cite{Townsend:1984iu,Boucher:1984yx} for two-derivative scalar theories with arbitrary potential, and reduces to the known condition as a particular case.
This result allows for more general $P$ than the recent result of~\cite{Nozawa:2013maa}, and we highlighted the technical step where our derivation deviates from theirs.

By examining a few special classes of $P$ we showed that in the $P(X)$ context it is possible to have positive energy while violating the DEC.
The derivation does however require that the NEC to be satisfied. More interestingly, it is possible to have positive energy
in cases where the S-matrix fails to satisfy the usual analyticity requirements for a local theory. It will be interesting to extend our results to more general derivative interactions, such as galileons or massive gravity~\cite{future}.

\noindent{\em Acknowledgments}: We thank L.~Berezhiani, R.~Deen, G.~Goon, M.~Nozawa,T.~Shiromizu and R.~Wald for helpful discussions and comments. This work is supported in part by NASA ATP grant NNX11AI95G (B.E. and J.K.) and NSF PHY-1145525 (J.K.); the Kavli Institute for Cosmological Physics at the University of Chicago through grant NSF PHY-1125897 and by the Robert R. McCormick Postdoctoral Fellowship (A.J.). A.J.T. is supported by DOE Early Career Award DE-SC0010600.



\begin{thebibliography}{99}
\bibitem{ftnote1}
For the ADM mass to be well-defined, we focus on flat (or AdS) asymptotics, where $\phi \rightarrow {\rm const.}$ at spatial infinity. This immediately rules out time-dependent asymptotics, $\phi \rightarrow \phi(t)$, which may be more realistic for cosmology.
\bibitem{ftnote2}
The DEC states that: $i)$ for any time-like $u$, $T_{\mu \nu} u^\mu u^\nu \geq 0$; and $ii)$ for any future-pointing and causal $u$, $-T^\mu_\nu u^\nu$ is also future-pointing and causal.
Roughly, these correspond respectively to the statements that the energy density is positive, and the energy-momentum flow is subluminal.
\bibitem{ftnote3}
In fact, $V(\phi^I)$ must only satisfy the weaker inequality $V(\phi^I) \geq 8f^{IJ}W_{,\phi^I} W_{,\phi^J}-12W^2$.
\bibitem{ftnote4}
The stress tensor for $P(X,\phi)$ is $T_{\mu\nu} = P_{,X}\partial_\mu\phi\partial_\nu\phi + g_{\mu\nu} P$. Contracting with
a null vector $n^\mu$, the NEC boils down to $0 \leq T_{\mu\nu} n^\mu n^\nu = P_{,X} (n^\mu\partial_\mu \phi)^2$, which requires
$P_{,X} > 0$.
\bibitem{ftnote5}
An alternative route to \eqref{PXmastereq} is to not introduce ${\cal G}_{,X}$ through~\eqref{spinordefs}, but demand that the second line of~\eqref{nesterdiv} (with ${\cal G}_{,X} = 0$) be positive definite. This yields the inequality $XP_{,X}-P-8{\cal W}^2_{,\phi}/P_{,X}+12{\cal W}^2 \geq 0$. Since this is positive-definite, we can write is as the square of some function. Calling this function ${\cal G}_{,X}$ yields~\eqref{PXmastereq}.
\end{thebibliography}
\end{document}